\documentclass[review]{elsarticle}

\usepackage{lineno,hyperref}
\modulolinenumbers[5]

\journal{Physics Letters B}

\usepackage{graphics} 
\usepackage{graphicx} 
\usepackage{epstopdf}
\usepackage{color}
\usepackage{scrtime}
\usepackage{amssymb}
\usepackage{amsmath}
\usepackage{amsfonts}
\usepackage{cleveref}
\usepackage[ulem=normalem]{changes}

\begin{document}

\begin{frontmatter}

\title{Nucleon-nucleon correlations and the single-particle strength in atomic nuclei}

\author[York]{S.~Paschalis\corref{mycorrespondingauthor}}
\cortext[mycorrespondingauthor]{Corresponding author}
\ead{stefanos.paschalis@york.ac.uk}
\author[York]{M.~Petri}
\author[Berkeley]{A.~O.~Macchiavelli}
\author[MIT]{O.~Hen}
\author[TelAviv]{E.~Piasetzky}

\address[York]{Department of Physics, University of York, York, YO10 5DD, UK}
\address[Berkeley]{Nuclear Science Division, Lawrence Berkeley National Laboratory, Berkeley, CA 94720,  USA }
\address[MIT]{Massachusetts Institute of Technology, Cambridge, MA 02139, USA}
\address[TelAviv]{School of Physics and Astronomy, Tel Aviv University, Tel Aviv, 69978, Israel}

\tnotetext[mytitlenote]{This article is registered under preprint number: 1812.08051}

\begin{abstract}
We propose a phenomenological approach to examine the role of short- and long-range nucleon-nucleon correlations in the quenching of single-particle strength in atomic nuclei and their evolution in asymmetric nuclei and neutron matter. These correlations are thought to be the reason for the quenching of spectroscopic factors observed in $\rm (e,e'p)$, $\rm (p,2p)$ and transfer reactions. We show that the recently observed increase of the high-momentum component of the protons in neutron-rich nuclei is consistent with the reduced proton spectroscopic factors. 
Our approach connects recent results on short-range correlations from high-energy electron scattering experiments with the quenching of spectroscopic factors and addresses for the first time quantitatively this intriguing question in nuclear physics, in particular regarding its isospin dependence. We also speculate about the nature of a {\sl quasi-proton} (nuclear polaron) in neutron matter and its
kinetic energy, an important quantity for the properties of neutron stars.
\end{abstract}


\end{frontmatter}


Many-body quantum mechanical systems consisting of interacting particles are encountered in many fields of modern physics, including condensed matter, atomic and nuclear physics. 
In general, it is not possible to obtain analytical solutions of the equations governing the dynamics of particles within such quantum systems, starting explicitly from the individual particle-particle interactions. Quantum Monte-Carlo and other numerical techniques can be used to obtain solutions, but these methods are computationally intense and are therefore limited to few interacting particles \cite{RevModPhys.87.1067, PhysRevC.89.024305, Lonardoni:2018prc}. To overcome these limitations, many-body systems are often described in terms of independent particles moving in an effective mean-field potential that reflects the average influence of all individual particle-particle interactions. 
In fermionic systems, neglecting any residual interactions between the particles (beyond those captured by the effective mean-field potential), one can define a Fermi level below which all quantum states are occupied. In the presence of residual interactions between fermions, important correlations arise that deplete the occupancy of states below the Fermi level and populate states above it, thus making the Fermi surface diffused.

The atomic nucleus consists of strongly interacting nucleons forming a dense quantum system. It is noteworthy that for such strongly interacting quantum system the independent-particle model (IPM) is proven to be a valid approximation and has provided the framework to explain many nuclear properties. Nevertheless, correlations between nucleons modify the mean-field approximation and dilute the pure independent-particle picture\footnote{Mottelson has given arguments based on the quantality parameter,  $\rm \Lambda = \frac{\hbar^2/M\alpha^2}{V_0}$, that nuclei should behave like a quantum Fermi fluid \cite{MOTTELSON199945}, with quasi-particles (qp) taking the role of the particles in the IPM. Where $\alpha$ is the inter-constituents distance.
 $\Lambda$ is a measure of the ratio of the zero point motion kinetic energy to the strength of the NN interaction.}. 
These nucleon-nucleon (NN) correlations are often distinguished into long-range correlations (LRC) and short-range correlations (SRC), referring to their spatial separation and the part of the NN potential they are most sensitive to \cite{DICKHOFF2004377,Hen:2016kwk,Frankfurt:1981mk}. Studies show that for stable nuclei at any given moment, only 60\% -- 70\% of the states below the Fermi momentum are occupied, with 30\% -- 40\% of the nucleons participating in LRC and SRC configurations \cite{DICKHOFF2004377, PhysRev.98.1445, BERTSCH1968204, LAPIKAS1993297, KRAMER2001267, Kay_PRL111_2013, PhysRevLett.110.122503, devins, Atkinson18}. Therefore, both LRC and SRC deplete the occupancy of single-particle states, with LRC primarily mixing states near the nuclear Fermi-momentum and SRC populating states well above it. 

There are two questions regarding this depletion that require further study, and have attracted the attention of the Nuclear Physics community:
\begin{itemize}
\item What are the individual contributions of LRC and SRC to the observed single-particle depletion?
\item What is the isospin (neutron-proton asymmetry) dependence of LRC and SRC, and how do they compete in very asymmetric nuclei?
\end{itemize}
Inspired by recent results from Jefferson Lab \cite{Duer_nature_18}, where the ratio of the fraction of high- to low-momentum protons (where high and low are relative to the Fermi momentum) was measured, we propose a phenomenological model that directly connects these new results with the reduction of single-particle strength in atomic nuclei. Our approach captures both LRC and SRC ingredients and allows one to extract their individual contributions as well as their evolution with mass number and isospin.

Experimentally, the depletion of single-particle states is quantified as quenching of spectroscopic factors (SFs) with respect to the IPM limit, observed in $\rm (e,e'p)$ \cite{LAPIKAS1993297, KRAMER2001267}, $\rm (p,2p)$ \cite{devins} and transfer reactions \cite{Kay_PRL111_2013, PhysRevLett.110.122503}. This is reflected in the probability to end up at a given final state after a nucleon is removed from the parent nucleus compared to theoretically calculated cross sections for the same reaction. At this point it is important to note that the quenching extracted from $\rm (e,e'p)$ measurements may depend on the momentum transfer, $\rm Q^2$ \cite{Lapikas:1999ss, Frankfurt:2000ty}. 
Although the $\rm Q^2$ dependence of the quenching  needs to be better understood, here we analyze the (well established) low-$\rm Q^2$ data, where the scale resolution should be sensitive to probe the quenching due to both SRC and LRC \cite{Lapikas:1999ss}.

Recently, single-nucleon removal \cite{gade, PhysRevC.90.057602} and hadron-induced (p,2p and p,pn) quasi-free scattering (QFS) reactions in inverse kinematics \cite{PhysRevLett.120.052501, Panin_PLB753_2016, doi:10.1093/ptep/pty011} have been employed using radioactive-ion beams and probed the quenching of SFs across a wider region of isospin asymmetry, exploring its isospin dependence. These experiments agree on the depletion of the single-particle strength for nuclei near stability but report significantly different isospin dependency. The reported discrepancy has triggered an active debate on the validity of the reaction models used in the analysis and the extent to which this can lead to an overestimation of isospin effects \cite{Kay_PRL111_2013, PhysRevLett.110.122503, gade, PhysRevC.90.057602, PhysRevLett.120.052501}.

In parallel, electron scattering experiments indicate a high-momentum tail extending far beyond the Fermi momentum \cite{RevModPhys.69.981, PhysRevLett.93.182501} attributed to SRC between a pair of strongly interacting nucleons \cite{Hen:2016kwk,Atti:2015eda}. A value of about 20\% SRC contribution was indirectly inferred from scaling inclusive measurements of the fraction of high-momentum nucleons in nuclei relative to deuterium \cite{Hen:2016kwk,Atti:2015eda,Fomin:2012,egiyan02,egiyan06,Frankfurt:1993sp}. 
Proton and electron scattering studies of $^{12}$C showed that SRC are predominantly neutron-proton (np) pairs, as opposed to proton-proton (pp) or neutron-neutron (nn) pairs that are favored at lower momenta \cite{piasetzky06,Subedi_science2008,shneor07}. This was interpreted as  manifestation of the tensor part of the NN interaction, which at short distances (q~$\approx$~2~fm$^{-1}$) favors the  $\rm S=1~(T=0)$ (quasi-deuteron) channel \cite{Hen:2016kwk,Atti:2015eda,Schiavilla:2006xx, Sargsian:2005ru, Alvioli:2007zz, Weiss:2016obx, Weiss:2018tbu}. Follow-up works extended these findings to both lighter and heavier nuclei, see e.g. Refs.~\cite{PhysRevLett.113.022501, Hen614, Duer:2018sxh}. 

Finally, the Jefferson Lab results from Ref.~\cite{Duer_nature_18} revealed that in neutron-rich nuclei the ratio of the fraction of high- to low-momentum protons increases as a linear function of the ratio of neutron to proton number (N/Z), while the equivalent fraction for neutrons is rather constant or possibly decreasing slightly. This indicates that the percentage of protons participating in SRC pairs increases for neutron-rich systems and consequently depletes the proton strength from the region below the Fermi momentum, which is probed in measurements of SFs. Hence the SRC dependence with isospin asymmetry should be reflected in the quenching of the proton SFs and furthermore
in other low-energy observables \cite{MILLER2019360}.

To study the consistency between SRC experimental results and SFs, we introduce a phenomenological model to estimate the total ``missing strength" in terms of contributions from LRC (defined here as pairing \cite{FRAUENDORF201424} and particle-vibration coupling) and SRC components.
While generally in low-energy nuclear structure one refers to
pairing correlations as the short-range part of the force, compared 
to the quadrupole force which is of longer range, within the context of 
this paper pairing is not part of the SRC associated with 
high-momentum components.
We approximate the wave function of a ``dressed'' nucleon (quasi-particle or qp) in the nuclear medium in the following form:
\begin{equation}
\rm |qp \rangle = K_{SP}|SP \rangle + K_{PVC}|PVC \rangle + K_{PC}|PC \rangle + K_{SRC}|SRC \rangle.
\label{eq1}
\end{equation}
The first term represents the pure single-particle (SP) configuration, and the following three terms the particle-vibration coupling (PVC), pairing correlations (PC) and SRC induced configurations, respectively. 
Thus, the overlap for the removal of a nucleon from the ground state of nucleus A to the ground state of  $\rm A-1$ is  $\rm \langle A-1 | a | A \rangle = K_{SP}$ and the probability to find a nucleon in the pure single-particle configuration is $\rm R$ = $\rm K_{SP}^2$. For non-interacting nucleons $\rm R = 1$, while in the presence of correlations $\rm R < 1$ (quenched). The missing part of the single-particle strength is distributed to the correlation terms with probabilities given by the square of the corresponding amplitudes in the wavefunction of \cref{eq1}, i.e. $\rm R_{PVC} = K_{PVC}^2$, $\rm R_{PC} = K_{PC}^2$ and $\rm R_{SRC} = K_{SRC}^2$. 
The quenched single-particle strength, R, can then be expressed in terms of these three independent components:
\begin{equation}
\rm R = 1- (R_{PVC}+R_{PC}+R_{SRC}).
\label{eq2}
\end{equation}
In this approach, we associate R to the overall quenching of SFs reported in $\rm (e,e'p)$ and $\rm (p,2p)$ measurements and extract the weighting of each of the three components entering \cref{eq2} as fitting parameters.

The linear form of \cref{eq1}, and lack of interference terms in \cref{eq2}, both stem from the underlying assumption that the SP, PVC, PC and SRC states are all orthogonal to each other. SRC induce mixing to states of very high momentum and energy in the nuclear spectral function and there should be a small overlap with the SP and LRC components, thus supporting the orthogonality assumption \cite{BERTSCH1968204,Atti:2015eda,Weiss:2018tbu}. 
Near doubly magic nuclei, for which both pairing and deformation manifest themselves as vibrations, the individual terms in \cref{eq1} can be justified in first order perturbation as one-particle-one-hole (1p1h) (PVC) and two-particle-two-hole (2p2h) (PC) excitations (see later discussions).

\begin{figure}
 \centering
  \includegraphics[width=0.8\textwidth, trim=0 210 0 210, clip=true]{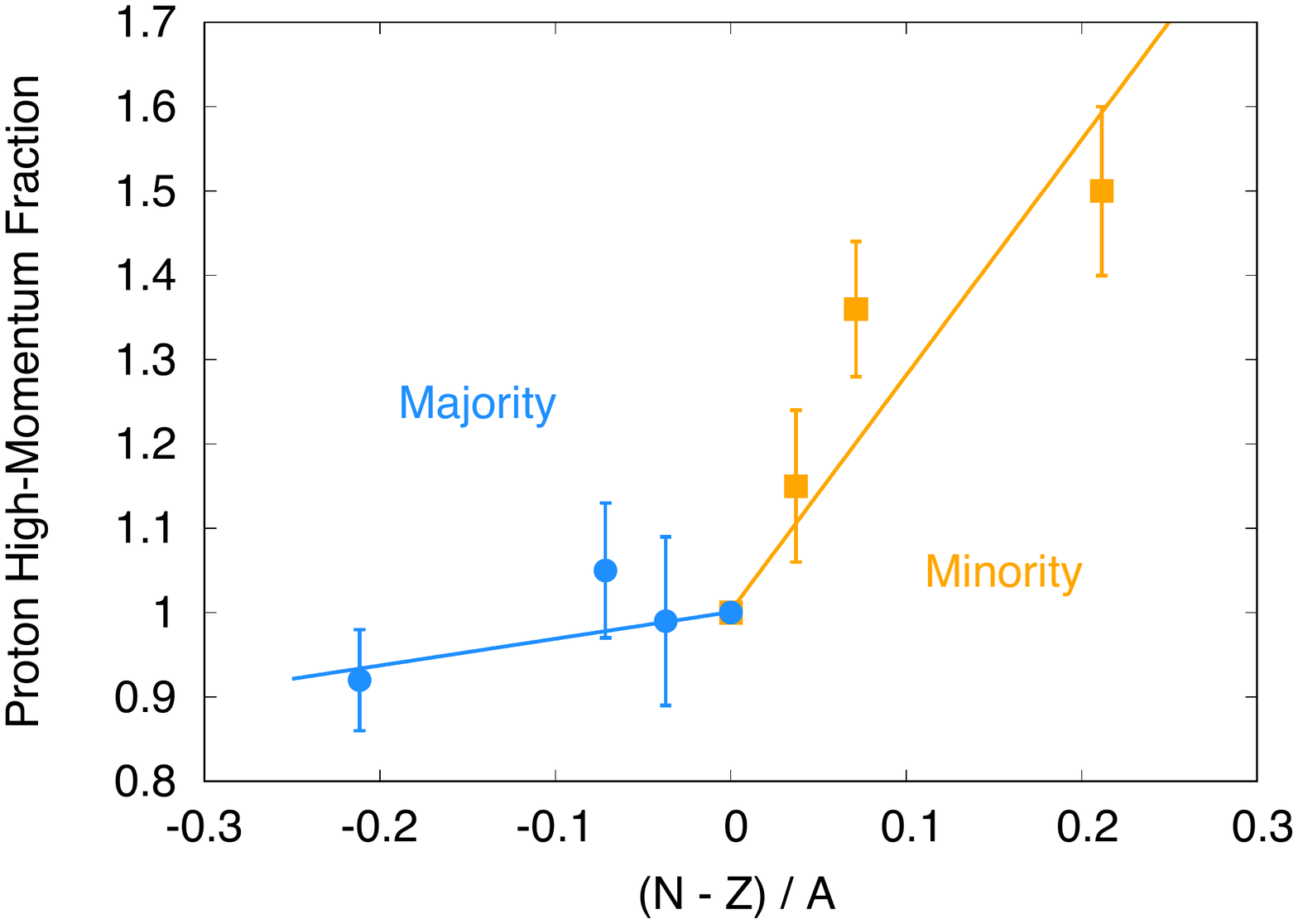}
 \caption{Proton high-momentum fraction relative to that of $^{12}$C as a function of $\rm (N-Z)/A$ and linear fits that give the slopes $\rm SL^p_{\text SRC} = 2.8\pm 0.7$ and $\rm SL^n_{\text SRC} = 0.3 \pm 0.2$. Note that the values for negative asymmetry correspond to the fraction measured for neutrons for positive asymmetry. Figure based on data from Ref.~\cite{Duer_nature_18}.}
 \label{Fig.1}
\end{figure}

The trend of the SRC component as a function of isospin is derived from Ref.~\cite{Duer_nature_18}.
The measured dependence of the relative fractions of high- to low-momentum nucleons in nuclei relative to $^{12}$C has been discussed in Ref.~\cite{RYCKEBUSCH201921}, where it was shown that
a low-order correlation operator approximation (LCA) can account for the 
observed trends in the data.
Here we reproduce the Jefferson Lab data in Fig.~\ref{Fig.1} (after transforming their N/Z axis to $\rm (N-Z)/A$) and make the assumption that the neutron momentum fraction measured for neutron-rich systems (neutrons being the majority nucleons) can be used as the proton momentum fraction in a proton-rich system (protons being the majority nucleons). Using the fitted slopes $\rm SL^p_{SRC} = 2.8\pm 0.7$ and $\rm SL^n_{SRC} = 0.3 \pm 0.2$, see Fig.~\ref{Fig.1}, we write the following expressions:
\begin{equation}
\rm N>Z : R_{SRC} =\gamma \bigg(1 + SL^p_{SRC} \frac{N-Z}{A}\bigg),
\label{eq3}
\end{equation}
 
\begin{equation}
\rm N<Z : R_{SRC} =\gamma \bigg(1 + SL^n_{SRC} \frac{N-Z}{A}\bigg).
\label{eq5}
\end{equation}

\begin{table}
\centering
\caption{SFs from $\rm (e,e'p)$ experiments \cite{KRAMER2001267,PhysRevC.73.044608} and their quenching, $\rm R = SF_{exp}/SF$, with respect to the SM, for ground-state to ground-state transitions. For doubly-magic nuclei (indicated with an asterisk in the last column), the SM SFs (and thus the overall quenching $\rm R$) are almost the same to the ones given by the IPM.}
\label{table:1}
\begin{tabular}{c|c|c|c}
\hline\hline
  Nucleus            & (N--Z)/A   & SF$_{\rm exp}$ &  R \\\hline
$^{7}$Li            & 0.143      & 0.42 $\pm$ 0.04     &                         0.63 $\pm$ 0.06    \\
$^{12}$C            & 0          & 1.72 $\pm$ 0.11     &                         0.60 $\pm$ 0.04    \\
$^{16}$O            & 0          & 1.27 $\pm $ 0.13    & 0.64 $\pm$ 0.07 *   \\
$^{30}$Si           & 0.067      & 2.21 $\pm$ 0.20      &                             0.58 $\pm$ 0.05    \\
$^{31}$P            & 0.032      & 0.40  $\pm$ 0.03     &                              0.69 $\pm$ 0.04    \\
$^{40}$Ca           & 0          & 2.58 $\pm $ 0.19      & 0.65 $\pm$ 0.05 *  \\
$^{48}$Ca           & 0.167      & 1.07 $ \pm $ 0.07     & 0.54 $\pm$ 0.04 *  \\
$^{51}$V            & 0.098      & 0.37 $\pm $ 0.03     &                           0.49 $\pm$ 0.04    \\
$^{90}$Zr           & 0.111      & 0.72 $\pm$ 0.07     &                              0.56 $\pm$ 0.05    \\
$^{208}$Pb          & 0.212      & 0.98 $\pm$ 0.09    & 0.49 $\pm$ 0.05 *  \\\hline\hline
\end{tabular}
\end{table}

The proton SF data used in this analysis are taken from $\rm A(e,e'p)$ experiments of Ref.~\cite{KRAMER2001267} and are summarized in Table~\ref{table:1}. The SFs include only reactions that have populated the ground state of the daughter nucleus. Also included in Table~\ref{table:1} is the quenching of SFs (R) with respect to large-scale shell-model (SM) calculations of Ref.~\cite{PhysRevC.73.044608}.
In SM calculations the reported SFs for doubly-magic nuclei is almost the same to that expected from an IPM picture (indicated in Table~\ref{table:1} with an asterisk). In other words, there is no quenching predicted by SM for these closed-shell systems. This is inconsistent with the experimentally reported values for the quenching of SFs for doubly-magic nuclei and reflects the fact that SM calculations cannot reproduce the full strength lost in LRC due to the yet limited model space used ~\cite{BERTSCH1968204, PhysRevLett.103.202502}, and the lack of the SRC component. 
Similarly, the reduced SFs obtained by the SM for non-doubly magic nuclei can also be regarded as additional LRC not completely captured by the SM even in a large model-space.

\begin{figure}
 \centering
 \includegraphics[width=0.8\textwidth, trim=0 210 0 210, clip=true]{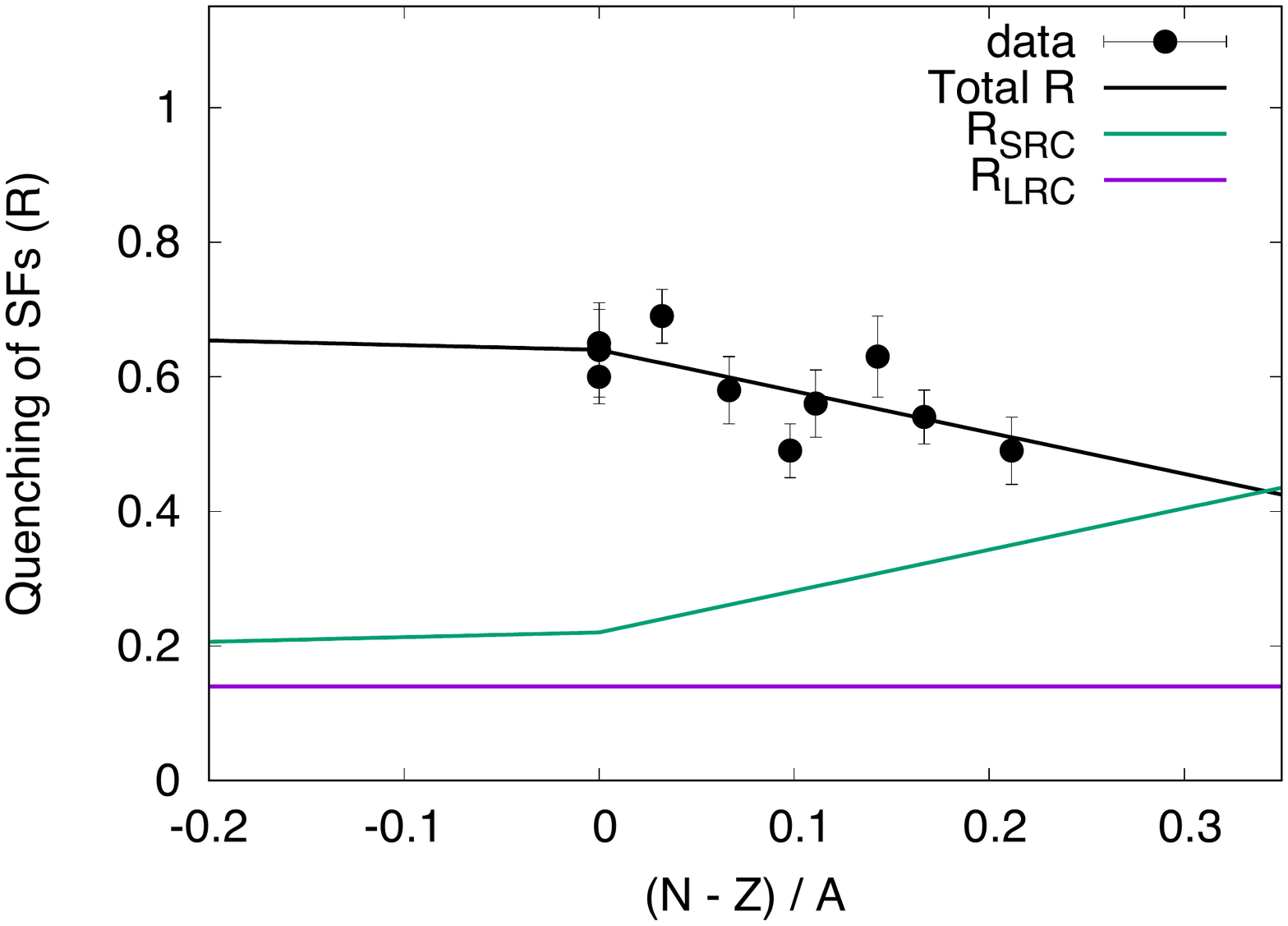}
 \caption{The quenching of proton SFs R as a function of $\rm (N-Z)/A$ for the $\rm (e,e'p)$ data shown in Table I. A fit to the data using \cref{eq2,eq3,eq5} is shown together with the individual SRC and LRC components.}
 \label{plot_kramer_all}
\end{figure}

Realizing that the available data are somewhat limited we make first the simplifying
assumption that $\rm R_{LRC} = R_{PVC} + R_{PC} = \delta$ is a constant as a function of isospin. We will further discuss this dependence later on.
While local deviations are expected, we believe that the overall constant trend (used in our first fit) is justified, as seen for example in
the Relativistic Mean Field
calculations of Ref.~\cite{PhysRevC.84.014305}. 
In Fig.~\ref{plot_kramer_all}, we show the results of the fit for the overall quenching of single-particle strength R (\cref{eq2}) together with the contribution by each of the individual components, as discussed above. To extract the errors on the individual components and investigate their statistical correlation, the fitting has been performed in a Monte-Carlo approach by re-sampling the values R within their probability density distribution (assumed to be Gaussian with standard deviation defined by their errors). The SRC contribution amounts to $\rm \gamma = 22\%~\pm~8\%$ and the LRC contribution to $\rm \delta = 14\%~\pm~10\%$. This is in accordance with expectations \cite{Hen:2016kwk, Duer_nature_18, Fomin:2012, Subedi_science2008,  Hen614}.  

While the simplest form introduced above ($\rm R_{LRC} = R_{PVC} + R_{PC} = \delta$) appears to capture the main ingredients of the quenching mechanism,  we now proceed to elaborate in more detail the individual contributions from PVC and PC with isospin dependence.

If due to PVC a single particle near a doubly-closed shell core is removed from its shell by coupling to surface phonons (built up from a coherent superposition of 1p1h excitations of multipolarity $\rm \lambda$), then the
single particle state $\rm |j_1\rangle$ becomes dressed and we write \cite{BrinkBroglia,BohrMottelsonVol-1}:
\begin{equation}
\rm { |\tilde j_1 \rangle} = m |j_1 \rangle +  \sum  \frac{ \langle j_2,n=1|H_c|j_1\rangle}{E_{j_2}-E_{j_1}+\hbar\omega_0} |j_2\rangle.
\label{eq:eq3}
\end{equation}
The quenching can be estimated by the amplitude of the coupling term, $\rm H_c$. 
Following Refs.~\cite{BrinkBroglia,BohrMottelsonVol-1}, this is proportional to the collectivity of the phonon (as measured by the dynamic deformation parameter $\rm \varepsilon_\lambda$) and the radial form factor, proportional to $\rm \partial V / \partial r$.
The potential depth (V) for a proton is often parametrized, including a term that depends on the neutron excess, as:
\begin{equation}
\rm V(r)=V_0(r) \bigg(1+\kappa \frac{N-Z}{A}\bigg).
 \label{eq7}
\end{equation}
We thus expect,
\begin{equation} 
\rm R_{PVC} \propto  \bigg(\frac{\varepsilon_\lambda}{\hbar \omega_0}\bigg)^2\bigg( \frac{\partial V}{\partial r}\bigg)^2.
\label{eq8}
\end{equation}
Using the potential given in Ref.~\cite{BohrMottelsonVol-1} (pg.~239, Eq.~2-182), we propose a parametrization of the form 
\begin{equation}
\rm R_{PVC} = \alpha \bigg(1+\frac{33}{51} \frac{N-Z}{A}\bigg)^2,
\label{eq9}
\end{equation}
with $\rm \alpha\propto (\varepsilon_\lambda / \hbar \omega_0)^2$.
For finite nuclei, $\rm \alpha$ can be taken as a constant, given the average dependence of $\rm \varepsilon_\lambda$ and $\rm  \hbar\omega_0$ with mass number. 
However, for infinite systems it should scale as $\rm 1/A^{1/3}$ reflecting the surface nature of the coupling.

In a similar way, we can estimate the effect of fragmentation due to pairing (vibration) correlations. The mixing amplitude should be proportional in lowest order to the ratio of the pairing gap ($\rm \Delta$) to a typical shell gap ($\rm \hbar \omega_0 = 41/A^{1/3}$~MeV), 
$\rm \Delta / \hbar \omega_0$. 
With the parametrization of 
$\rm \Delta$ in Ref.~\cite{JENSEN1984393} 
we obtain: 
\begin{equation}
\rm R_{PC} \propto \bigg(\frac{\Delta}{\hbar \omega_0}\bigg)^2 = \beta \bigg(1-6.07 \bigg(\frac{N-Z}{A}\bigg)^2\bigg)^2,
\label{eq10}
\end{equation}
here, $\rm \beta$ is also a constant. Note that, specifically for $\rm N=Z$, we have $\rm \delta=\alpha+\beta$.
Turning again our attention to the doubly magic nuclei, for which to lowest order pairing vibrations will 
introduce 2p2h admixtures in the unperturbed (0p0h) ground-state configuration, 
one can make a simple estimate of $\rm \beta$ as $\rm ((7.55/A^{1/3})/(41/A^{1/3}))^2 \approx 0.03$, 
using specifically $\rm \Delta_p = 7.55(1-6.07(N-Z)^2A^{-2})/A^{1/3}$~MeV \cite{JENSEN1984393}, since we are looking at proton spectroscopic factors.

\begin{figure}
 \centering
 \includegraphics[width=0.8\textwidth, trim=0 210 0 210, clip=true]{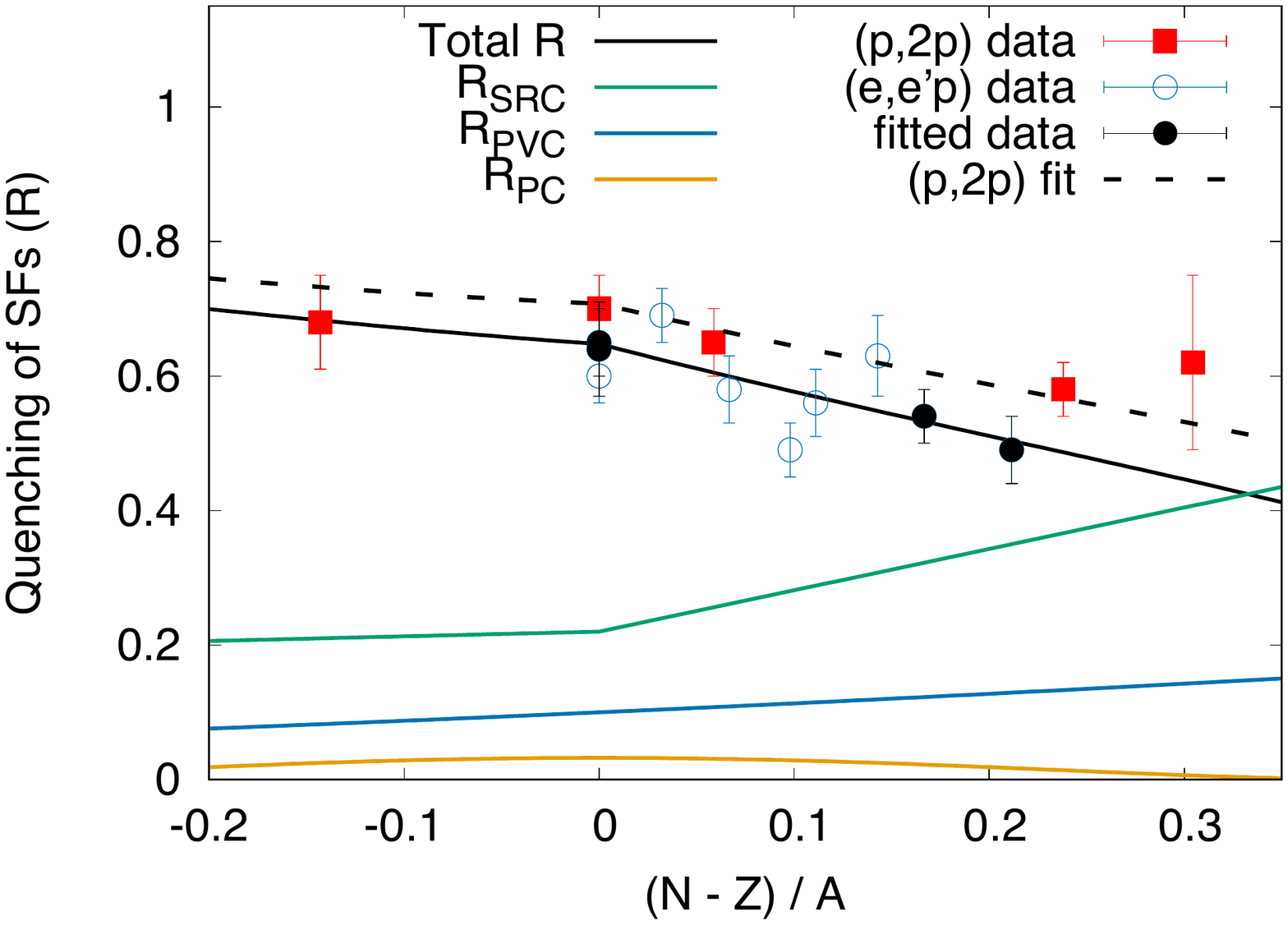}
 \caption{The full set of $\rm (e,e'p)$ data and $\rm (p,2p)$ results from Ref.~\cite{PhysRevLett.120.052501}. As discussed in the text, the fit corresponds to doubly magic nuclei only. For comparison, the dashed line shows the 
fit for the $\rm (p,2p)$ data. The SRC and PC contributions are fixed to $\rm \gamma = 22\%$ and $\rm \beta=3\%$, respectively. The fit yields a PVC contribution of $\rm \alpha=10\% \pm 2\%$ for ground- to ground-state transitions and a smaller PVC contribution of $\rm \alpha=4\% \pm 2\%$ for the 
$\rm (p,2p)$ results from Ref.~\cite{PhysRevLett.120.052501}; this is expected since the (p,2p) data is an inclusive measurement.}
 \label{R_vs_NZA_doublyMagic_plot_all}
\end{figure}

With the expressions in \cref{eq2,eq3,eq5,eq9,eq10} we attempt a fit of the experimental data on doubly magic nuclei. The result of the fit gives a PVC contribution of 
$\rm \alpha=10\%~\pm~2\%$. The SRC and PC contribution have been fixed to $\rm \gamma = 22\%$ (from the fit of Fig.~\ref{plot_kramer_all}) and $\rm \beta = 3\%$, respectively, based on the above arguments. The total fit (and the individual components) shown in Fig.~\ref{R_vs_NZA_doublyMagic_plot_all} is in good agreement with the full $\rm (e,e'p)$ data set.
As discussed earlier, the agreement seen also for open-shell nuclei indicates a level of missing strength in the full shell-model results, due to LRC, similar to that in the IPM for doubly-magic systems.
QFS $\rm ^{A}O(p,2p)$ data \cite{PhysRevLett.120.052501} are also shown in the same plot. The reported QFS data are inclusive measurements to all bound final states and not only ground-state to ground-state transitions like the $\rm (e,e'p)$ data. This means that part of the LRC correlations that distributes the single-particle strength to low-lying excitations in the final states is integrated in the experimental cross section. Indeed, repeating the fit for the  $\rm ^{A}O(p,2p)$ data, we  obtain a PVC contribution of $\rm \alpha=4\%~\pm~2\%$. Actually, the $\rm ^{A}O(p,2p)$ measurement has the only data point at negative asymmetry, which is nicely reproduced with the different slope of the SRC contribution for removing a proton from a proton-rich system (see \cref{eq5}). In this case, the protons are the majority nucleons. 

\begin{figure}
 \centering
 \includegraphics[width=0.7\textwidth, trim=10 0 50 20, clip=true]{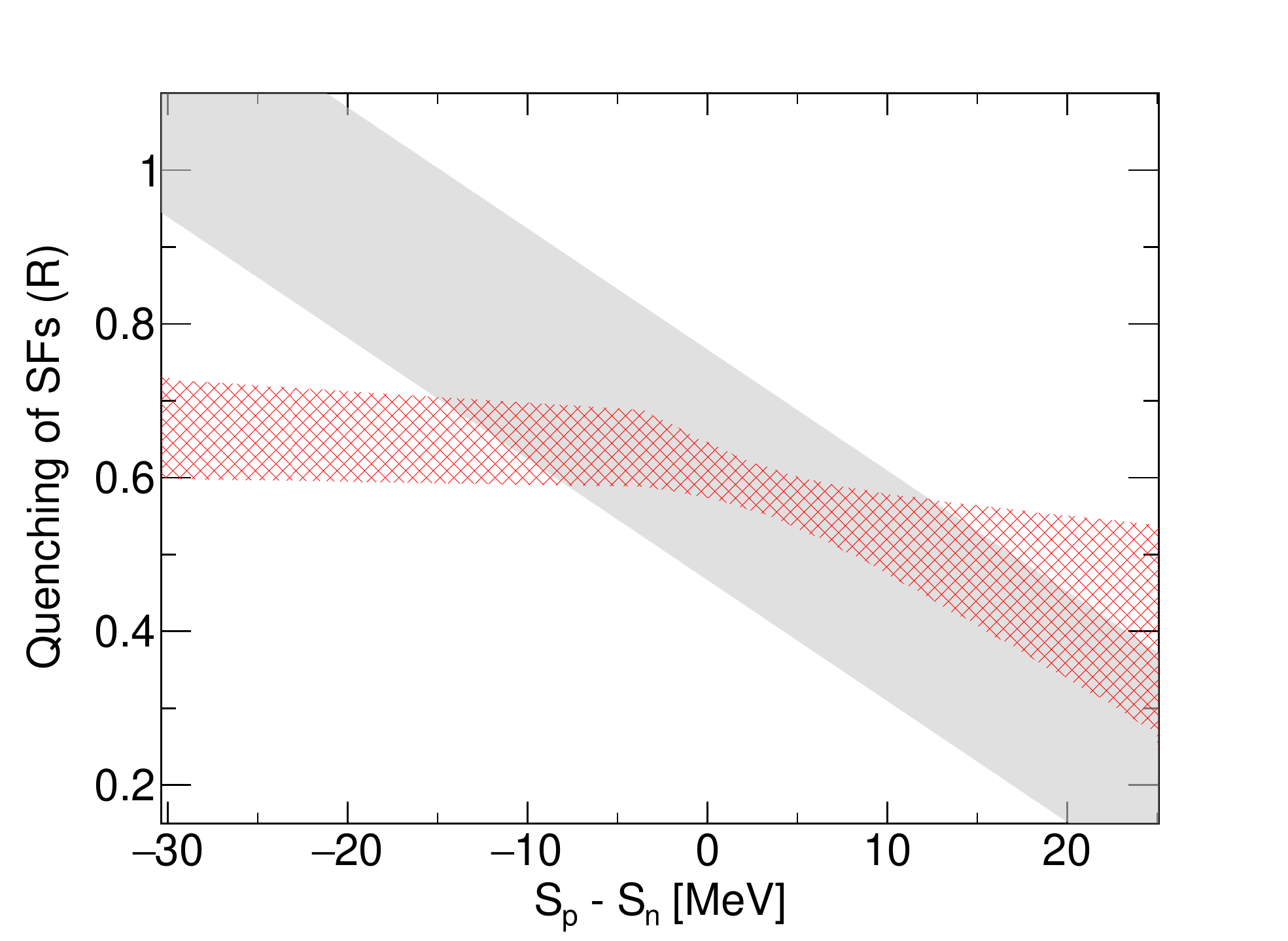}
\caption{The grey-shaded area shows the quenching of proton SFs measured in nucleon-removal reactions \cite{PhysRevC.90.057602} as a function of the difference in separation energies $\rm S_p - S_n$. Our prediction (within $\rm 2 \sigma$) follows the red-shaded area.  }
\label{gadePlot}
\end{figure}

Another topic of current debate is the quenching observed in one-proton (and one-neutron) removal reactions carried out at intermediate energies 
($\rm \sim\,100\,MeV/nucleon$). The study of Ref.~\cite{PhysRevC.90.057602} showed an unexpectedly strong dependence of the quenching, expressed as a function of the difference ($\rm \Delta S$) in proton and neutron separation energies, $\rm S_{p}-S_{n}$ ( $\rm S_{n}-S_{p}$).
The origin of this strong dependence (i.e. whether it is indeed due to NN correlations or due to the reaction model) is still an open question.  
To add to the discussion,  it is interesting to compare our predictions (from Fig.~\ref{plot_kramer_all}) with the results of Ref.~\cite{PhysRevC.90.057602}.  For this purpose, we use the equations given in Ref.~\cite{VOGT2001255} to convert  A, Z, and N into $\rm S_{p}-S_{n}$.  For the conversion from $\rm (N-Z)/A$ to $\rm \Delta S = S_{p}-S_{n}$ we derive the equation  $\rm \Delta S = (75\rm (N-Z)/A - 3.4)~\rm MeV$, which holds for light and medium mass nuclei,
considered in the systematics of Ref.~\cite{PhysRevC.90.057602}. 
The two trends are shown
as shaded areas in Fig.~\ref{gadePlot}.  Our results give a less pronounced dependence on $\rm \Delta S$, 
suggesting that the theory of the reaction mechanism is playing a role in the measured $\rm \Delta S$ dependence of
Ref.~\cite{PhysRevC.90.057602}.

As a final note, we can also speculate about the nature of a {\sl quasi-proton} (nuclear polaron \cite{PhysRevC.47.1077})
in neutron matter (nM).  In the limit of $\rm A \rightarrow \infty$ and $\rm (N-Z)/A \rightarrow 1$, and neglecting both surface and pairing coupling terms, expected to be small for infinite matter at saturation density, we predict
a proton quenching factor of $\rm R_{nM}^p = 1- \gamma-\gamma SL^p_{\text SRC} \sim 0.16$.
The high relative kinetic energy components in the wavefunction, present in this limit, will give the proton an average kinetic energy:
\begin{equation}
\rm \Big \langle T_p \Big \rangle_{nM} = \Big ( R_{nM}^p+ \Big (1-R_{nM}^p\Big) \frac{5}{3}\frac{p_{Max}}{p_F}\Big ) \Big \langle E_F \Big \rangle,
\label{eq11}
\end{equation}
from the nucleon momentum distribution in nuclear matter. 
If we assume $\rm p_{Max} \sim 2p_{F}$,
above which the two-nucleon wave function that dominates the SRC tail decays very rapidly \cite{Hen:2016kwk,Weiss:2018tbu, WEISS2018211, PhysRevC.91.025803, PhysRevC.92.045205,  YONG2017104}, 
we obtain $\rm \big \langle T_p \big \rangle _{nM}$ of approximately 3 times 
that of a proton in a Fermi Gas, an important quantity for the properties of neutron stars \cite{YONG2018447}.
Similarly, the average kinetic energy for the neutrons $\rm \big \langle T_n \big \rangle _{nM}$ (for which $\rm R_{nM}^n = 1-\gamma+\gamma SL^n_{\text SRC} \sim 0.85$) is estimated to be approximately 1.4 times 
that of a neutron in a Fermi Gas.

In summary, we presented a phenomenological model that connects the quenching of spectroscopic factors with the recent SRC Jefferson Lab study.  We derived simple phenomenological parametrizations for the combined effects of SRC, PVC, and PC that were used in an analysis of  data from low-$\rm Q^2$ electron scattering and proton induced QFS experiments. Our analysis shows that approximately 20\% of the missing strength observed in the region of $\rm N = Z$ can be attributed to SRC, in agreement with reported expectations. 
Furthermore, we
show how the missing strength, including contributions from LRC, is expected to evolve with $\rm (N-Z)/A$ and speculate an extrapolation to a {\sl quasi-proton} in neutron matter and its kinetic energy, with implications to neutron stars.
While perhaps rather speculative at this stage, given the available data, we trust our conjecture will stimulate further theoretical and experimental work. In particular for the latter, we highlight the need to:  
1) Measure the quenching of spectroscopic factors in stable nuclei with higher precision, including transfer reactions,
2) Study of $\rm (p,2p)$ reactions on very asymmetric nuclei available at radioactive beam facilities, and 
3) Extend those studies to neutron removal reactions.

\section*{Acknowledgements}
This work is supported by 
the UK STFC awards ST/M006433/1 and ST/P003885/1,
by the Royal Society award UF150476,  by the U.S. Department of Energy, Office of Nuclear
Physics, under contract nos DE-AC02-05CH11231 (LBNL) and DE-FG02-94ER40818, by the Israel Science Foundation and the Pazy Foundation.
Enlightening discussions with R. Casten, J. Dobaczewski, R. Liotta, A. Pastore, and J. Schiffer are gratefully acknowledged.  

\section*{References}


\end{document}